\documentclass[numbib]{comnet-arxiv}
\usepackage[square,numbers,sort&compress]{natbib}
\usepackage{graphicx}
\usepackage{booktabs} 
\usepackage[utf8x]{inputenc} 
\usepackage{amssymb}
\usepackage{subcaption}
\usepackage{tikz}
\usepackage{xspace}
\usepackage{url}
\usepackage{array}

\newcommand{\xhdr}[1]{\paragraph*{\bf #1}}

\newcommand\clap[1]{\hbox to0pt{\hss#1\hss}}
\newcommand\MC[1]{%
        \multicolumn{2}{c}{\clap{\textbf{#1}}}%
}

\usepackage{cleveref}
\renewcommand{\tabcolsep}{0.2cm}

\graphicspath{{figs/}}

\jno{xxxxxx} 
\received{XX XX XXXX}
\revised{XX XX XXXX}
\accepted{XX XX XXXX}

\begin{document}

\title{Benchmarking community detection methods on social media data}
\author{Conrad Lee
\address{Clique Research Cluster\\
         University College Dublin\\
         8 Belfield Office Park, Clonskeagh\\
         Dublin 4, Ireland}
\email{conradlee@gmail.com}
\and{P\'{a}draig Cunningham}
\address{Clique Research Cluster\\
         University College Dublin\\
         8 Belfield Office Park, Clonskeagh\\
         Dublin 4, Ireland}
\email{padraig.cunningham@ucd.ie}
}
\maketitle

\begin{abstract}
  {Benchmarking the performance of community detection methods on
    empirical social network data has been identified as critical for
    improving these methods. In particular, while most current research
    focuses on detecting communities in data that has been digitally
    extracted from large social media and telecommunications services,
    most evaluation of this research is based on small, hand-curated
    datasets. We argue that these two types of networks differ
    so significantly that by evaluating algorithms solely on the
    former, we know little about how well they perform on the
    latter. To address this problem, we consider the difficulties that
    arise in constructing benchmarks based on digitally extracted
    network data, and propose a task-based strategy which we feel
    addresses these difficulties. To demonstrate that our scheme is
    effective, we use it to carry out a substantial benchmark based on
    Facebook data. The benchmark reveals that some of the most popular
    algorithms fail to detect fine-grained community structure.}
  {community detection, benchmarking, evaluation, social networks,
    datamining, social media data}
\end{abstract}
\section{Introduction}
\label{sec:intro}
Community structure has been identified as playing a key role in the
formation and function of many systems, so it
comes as no surprise that the topic topic has received a large amount
of attention, with hundreds of papers currently published on the topic
every year. However, in the literature on the topic it is
commonly observed that although many community detection methods
exist, we do not know which ones work best on real data
\citep{Fortunato2010,Lancichinetti2010,Yang2012a,Yang2012b}.

This problem of evaluation on real data is so acute that, in what has
been recognized as the authoritative review on the community detection
problem, Fortunato states that our inability to properly
benchmark algorithms has led to ``a serious limit of the field'' and
that little is known about which methods perform best in practice \citep{Fortunato2010}.

In this paper, we focus on this problem of evaluating community
detection methods, focusing on social networks. We first briefly
review the history of community detection in order to reveal why the
benchmarking and evaluation are currently so problematic. Next we
argue that recent attempts to solve the problem of evaluation using
large network datasets with ``ground truth'' data---while a step in
the right direction---are flawed because they do not properly deal
with the fact that the ground truth data is imperfect and
incomplete.

We propose a modified benchmarking workflow which we believe
appropriately utilizes an imperfect set of ground truth
communities. In this bencmark, we use the communities detected by an
algorithm to infer the value of node attributes related
to community structure. The inference is done in a machine-learning
setting where the community assignment matrix is used
as the features associated with each node. The idea is that if the
community detection algorithm has done a good job, then a machine
learning classifier should be able to use the community assignment
matrix to accurately inferring such an attribute.
Finally, we run a benchmark using our proposed workflow on forty
Facebook datasets. We find that none of the methods we test
automatically detect communities at all scales, and the methods become
most effective when run many times, each time with a different value
for the resolution parameter.


\section{A Brief History of Community Detection}
\label{sec:history}
We present the following brief history of community detection because
we believe the historical trends we highlight are closely related to
the currently inadequate standard of evaluation in the field of
community detection. In short, we argue that up until the mid-1990s, community
detection algorithms were typically run on smaller,
\emph{hand-curated} datasets that were gathered specifically for
research. Researchers typically carefully curated these datasets and
had well-informed prior knowledge on the community structure.  This
expert knowledge, based on first-hand experience with the social
system from which the data was generated, could be used to set up
benchmarks for community detection algorithms. However, since then the
focus has shifted to finding communities in larger datasets that are
not purpose gathered, but rather \emph{digitally extracted}, i.e.,
mined from sources such as log files of web services and mobile
communication networks. While the community structure found in these
datasets is potentially very different, researchers continue to
evaluate their algorithms on the smaller, purpose-gathered
datasets. This leaves us ignorant of how well community detection
algorithms perform on larger, mined datasets.

\begin{figure}[t]
  \centering
  \begin{subfigure}[]{0.45\linewidth}
    \includegraphics[width=0.9\linewidth]{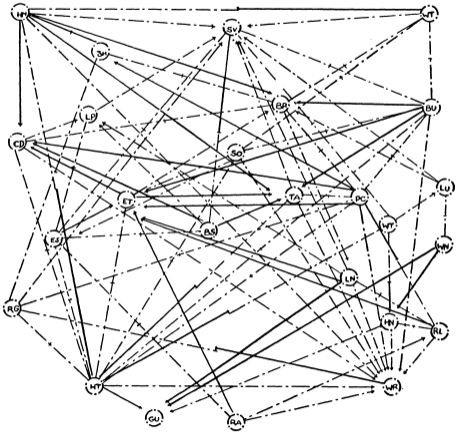}
    \caption{One of Moreno's early sociograms}
    \label{fig:sociogram}
  \end{subfigure}~\hspace{0.5cm}
  \begin{subfigure}[]{0.45\linewidth}
    \centering
    \includegraphics[width=0.95\linewidth]{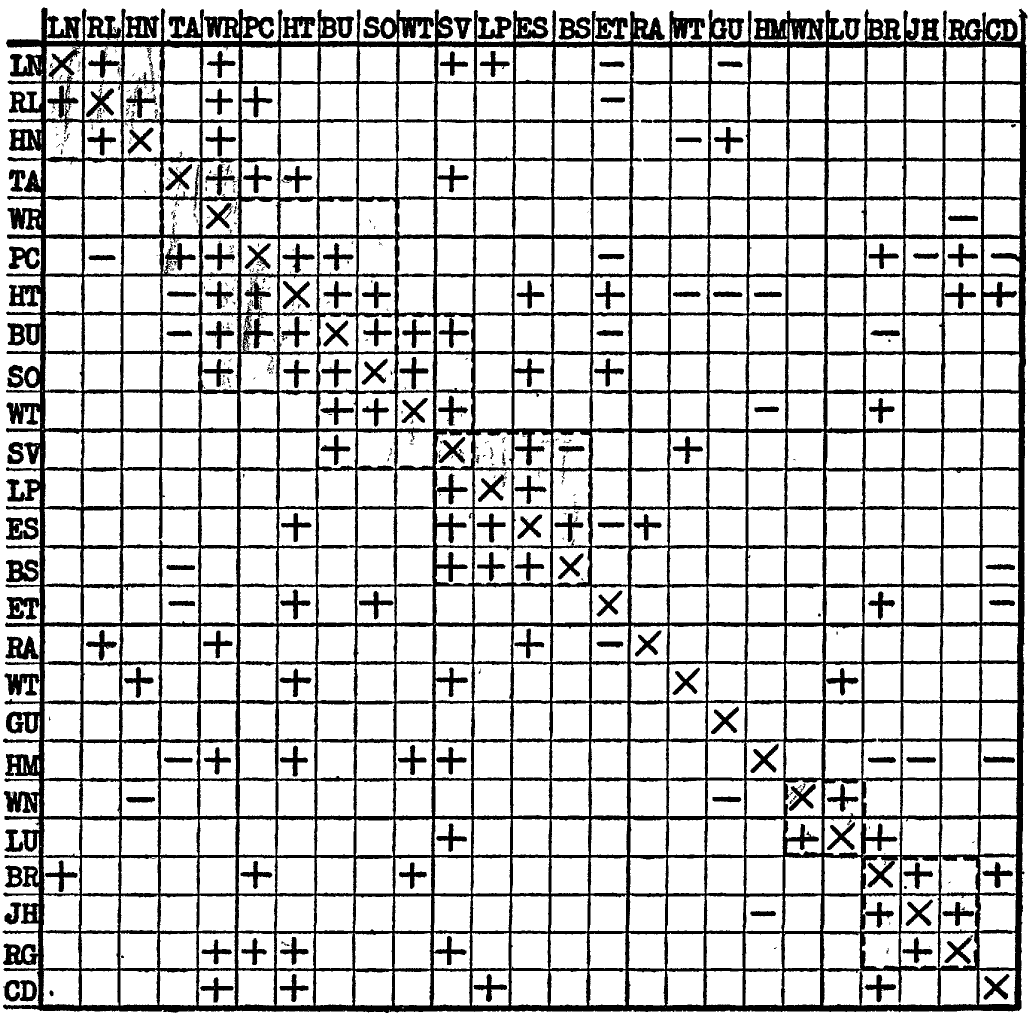}
    \caption{The same network depicted using Forsyth et al.'s
      community detection method.}
    \label{fig:sociomatrix}
  \end{subfigure}
  \caption{The first community detection algorithm, introduced by
    Forsyth et al. in 1946 was supposed to more objectively and
    clearly display community structure \citep{Forsyth1946}. Because the network used for
    evaluating the utility of the technique is small and hand-curated,
    the evaluation was informal, yet adequate.}
  \label{fig:historical}
\end{figure}

With the objective of examining how improvements in community
detection methods were evaluated, let us start by examining one of the
earliest improvements: the introduction of the adjacency matrix as a
means of group detection. The adjacency matrix was meant to improve
upon the ``sociagram'' introduced by Moreno in 1934
\citep{Moreno1934}. An example of a sociogram is depicted in
\cref{fig:sociogram}. While the sociogram as a method for
visualization attracted immediate attention (the New York Times
described it as a ``new science'' \citep{Unknown1933}), laying out
nodes and links by hand was criticized as too subjective: ``at
present, the sociogram must be built by a process of trial and error,
which produces the unhappy result that different investigators using
the same data build as many different sociograms as there are
investigators'' \citep{Forsyth1946}. If one looks at
\cref{fig:sociogram}, one can observe the grounds for this criticism:
not only is it time-consuming to draw such a diagram by hand, but it
also seems that the group structure of the social network is not very
clear, and that it might be clearer if one had drawn it differently.

In 1946, Forsyth et al. proposed one could represent social networks
with an adjacency matrix \citep{Forsyth1946}. The focus of their paper
was a procedure for sorting the rows and columns of this matrix in
order ``to present sociometric data more objectively, and to make
possible a more detailed analysis of group structure.'' One can argue
that the method presented in that paper is the first community
detection algorithm.  The algorithm orders the entries in an adjacency
matrix such that the community structure should be apparent as dense
blocks along the diagonal.  In \cref{fig:sociomatrix}, we see the same
social network as in \cref{fig:sociogram}, but this time visualized
using the technique proposed by Forsyth et al.

For our purposes, the most relevant part of this paper is how
Forsyth et al. evaluate how well their method works. Their evaluation
can best be described as a sort of informal visual check. They present
\cref{fig:sociomatrix} and assert that the group structure is much clearer than
it was when depicted using Moreno's sociogram method. They outline what they
consider the groups to be in \cref{fig:sociomatrix} using faint dotted lines to
draw blocks along the diagonal axis. They simply assert that these boxes represent
subgroups, even though these subgroups are not described by Moreno. The important
point is that, aside from the visualization, there is no external evidence to support
their claim that the subgroups are valid. Thus, the evaluation employed here
boils down to visualizing the output of the algorithm and asserting that one can
clearly see group structure.

In one sense, this is a poor evaluation: aside from their informal
visual check, Forsyth et al. present basically no empirical
evidence that their method more objectively
detects community structure. A proper empirical evaluation might
begin with several networks where the ``ground truth'' set of
communities is known, and then determine whether users of the new
method are able to more accurately identify the ground truth set of
communities than when they use previously-existing techniques.

However, from another perspective, the evaluation is adequate. The
researcher who gathered data such as in the example above (in this
case H. H. Jennings) typically spends months carrying out surveys on
and observing the social system which produces the data. Even before
any social network analysis is performed, the researcher has a rich
understanding of the social structure that exists among the
subjects. In this context, community detection methods are not meant
to discover previously unknown structure; rather, they are meant to
support, augment, and ``make objective'' the expert knowledge built up
over months of observation and first-hand research.

Over the next fifty years, the datasets used to evaluate community
detection algorithms were also generally gathered by experts who had
first-hand knowledge of the social system which the dataset
covered. Well-known examples include Zachary's
Karate Club \citep{Zachary1977}, Sampson's Monks \citep{Sampson1968},
and the Southern Women dataset \citep{Davis1941}. Through their close
observation, the researchers who collected this data were able to
group the nodes into communities based on events such as crises or
social gatherings. During this time, network datasets tended to be
small (with fewer than 500 nodes, and often fewer than 50 nodes) and
well-studied (in \citep{Freeman03}, Freeman synthesizes
the findings of 21 methodological studies on the Southern Women's
network alone).

Then, starting in the late 1990s, a new era of work on community
detection began. This new era was created in part by a new type of
social network data that emerged in the form of digital records such as
mobile communication records or data from Facebook. While this new
data still represents social networks, it differs from the data that
had be analyzed in the previous decades in important ways because it
is not collected personally by researchers.

\begin{figure}[t]
  \centering
  \begin{subfigure}[]{0.45\linewidth}
    \includegraphics[width=0.9\linewidth]{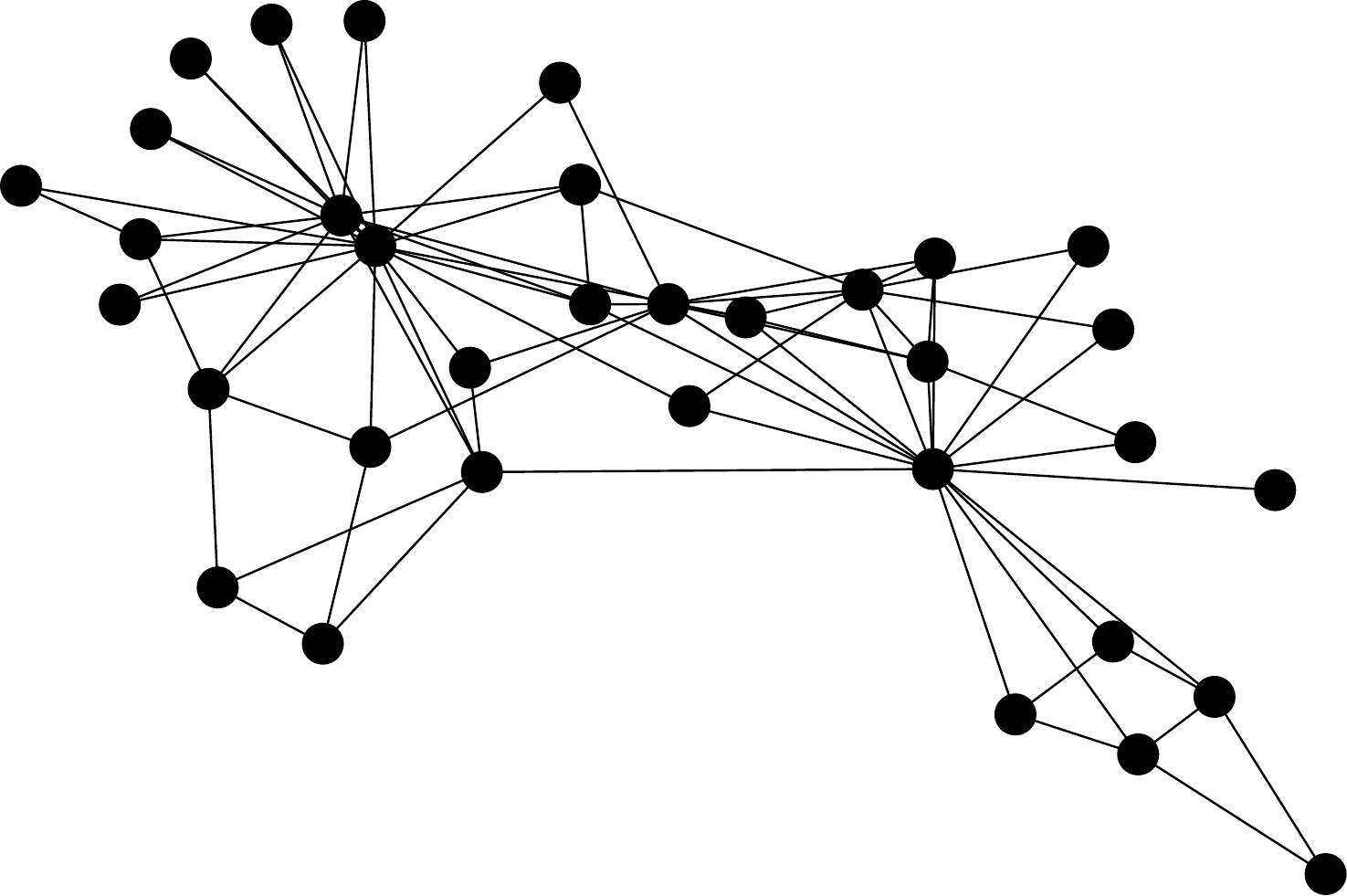}
    \caption{Zachary's Karate Club \citep{Zachary1977}}
    \label{fig:zachary}
  \end{subfigure}~
  \begin{subfigure}[]{0.45\linewidth}
    \includegraphics[width=0.72\linewidth]{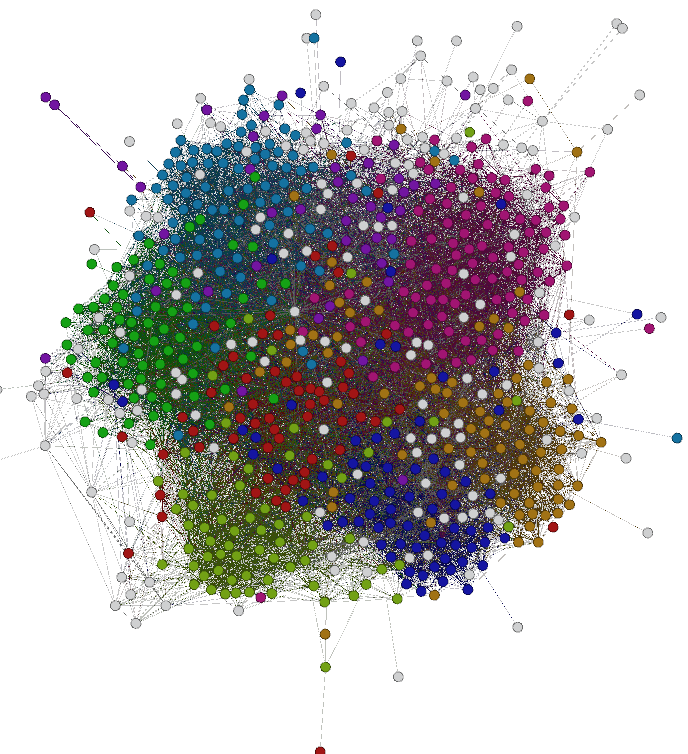}
    \caption{Facebook friendships at Caltech  \citep{Traud2012}}
    \label{fig:facebook-caltech}
  \end{subfigure}
  \caption{On the left, a network which is
    typical of ``hand curated'' datasets gathered by a researcher in
    the field. On the right an example of a ``digitally extracted''
    dataset. We have reason to believe that the structure of the
    communities in these two types of networks differs in important ways.}
  \label{fig:data-contrast}
\end{figure}

First, the new data sets are typically not collected specifically for
the purpose of a scientific study, but rather extracted after the fact
from logs or databases. As a result, the data may be messier (e.g., include a
majority of users who have very low activity levels) and cover many
social contexts for each user. For example, in most of the datasets
gathered up to the mid-1990s, only one social context would be
studied, such as activity in a club, at home, or in the workplace. On
the other hand, in the more modern datasets, such as Facebook data,
social interactions from several social contexts are jumbled
together. Another important difference is that the more modern
datasets are typically several orders of magnitude larger than the
earlier type. \Cref{fig:data-contrast} displays an example of what we
will call a \emph{hand-curated network} and a \emph{digitally extracted network}.
Finally, whereas the goal of community detection on hand-curated
networks was often to make essentially known community structure more
objective, the goal on large digitally extracted networks is to
uncover completely unknown community structure.

Due to these key differences between the old, hand-curated networks
and the new, digitally extracted networks, many new methods for the
community detection problem were proposed. The field of community
detection enjoyed booming popularity as more physicists, computer
scientists, and social scientists developed these new methods.  We
argue that when modern community detection methods were evaluated on
social network data something went wrong: rather than evaluating these
new methods on the new datasets for which they were designed, the new
methods were often evaluated on the old
datasets. \citep{Girvan2002,Newman2006,Clauset2007,Duch2005}\footnote{In
  some of these papers a larger social network was evaluated (such as
  the co-authorship network on arXiv), but these lacked the
  ground-truth or meta-data necessary for a proper evaluation. These
  larger networks were typically employed only for comparing something
  other than how well the algorithm identifies all relevant community
  structure, such as which algorithm gets the highest modularity or
  runs quicketst.} Thus, we know that many of the new community
detection methods work well on datasets like Zachary's Karate Club or
the Southern Women's dataset, but we do not know how well they work on
larger, digitally extracted datasets, and this is the ignorance that
Fortunato described in the excerpt above.

Note that these new methods were evaluated on diverse types of data.
For example, the Gene Ontology and other annotation can be used to
evaluate the modules found in protein-protein interaction networks
\citep{Marras2010}; product categorizations can be used to annotate the
network of products co-purchased on Amazon.com \citep{Yang2012b}. See
\citep{Ahn2010} for an example of thorough evaluation on current
datasets sets from these and other types of data.  Here we consider
evaluation only on social network data, i.e., networks in which nodes
represent humans and links represent relationships.

There is thus a clear need for benchmarks based on modern, digitally
extracted social network datasets, however creating these is not
straightforward. Researchers who gather a small dataset by hand for
the purpose of studying group behavior can confidently annotate their
datasets with the ground truth community structure. On the other hand,
with the digitally extracted datasets it is nearly always unclear how
exactly one should define a ground-truth set of communities. In some
cases, it might not even be clear that such community structure exists
in the data at all.

If we are to make progress in detecting community structure in modern
datasets, then it is imperative that we design benchmarks based on
such data. In the next section, we review recent efforts in this
direction and propose how such a benchmark could be created.


\section{Evaluation on digitally extracted networks}
\label{sec:annotated}

Recent efforts to benchmark the performance of community detection
algorithms can broadly be placed into three categories:
\begin{enumerate}
\item \emph{real-world benchmarks}, such as Zachary's Karate Club, where a
  dataset based on a social system includes a natural set of
  ground-truth communities;
\item \emph{synthetic benchmarks}, where data is artificially created
  according to some model which includes a pre-defined set of
  ground-truth communities; and
\item \emph{task-oriented benchmarks}, in which communities are used to help
  complete some task on real-world data, such as graph compression,
  decentralized routing \citep{Stabeler2011}, or attribute inference \citep{Mislove2010}.
\end{enumerate}
The real-world benchmark is ideal, because clear problems exist
with the other options. While using synthetic data to
benchmark an algorithm is better than nothing, it is unclear how to
create synthetic data that is realistic. Thus, even if an algorithm
performs well on synthetic data, it may perform poorly on real datasets.
The problem with task-oriented benchmarks is that many types of
structure may be useful for solving a task, including structures that
do not resemble network communities. Thus, the algorithm that performs
best on the task-oriented benchmark may not be finding communities at
all, but something else, e.g., the clusters identified by block
modelling.

\begin{figure}[h]
  \centering
  \includegraphics[width=0.65\textwidth]{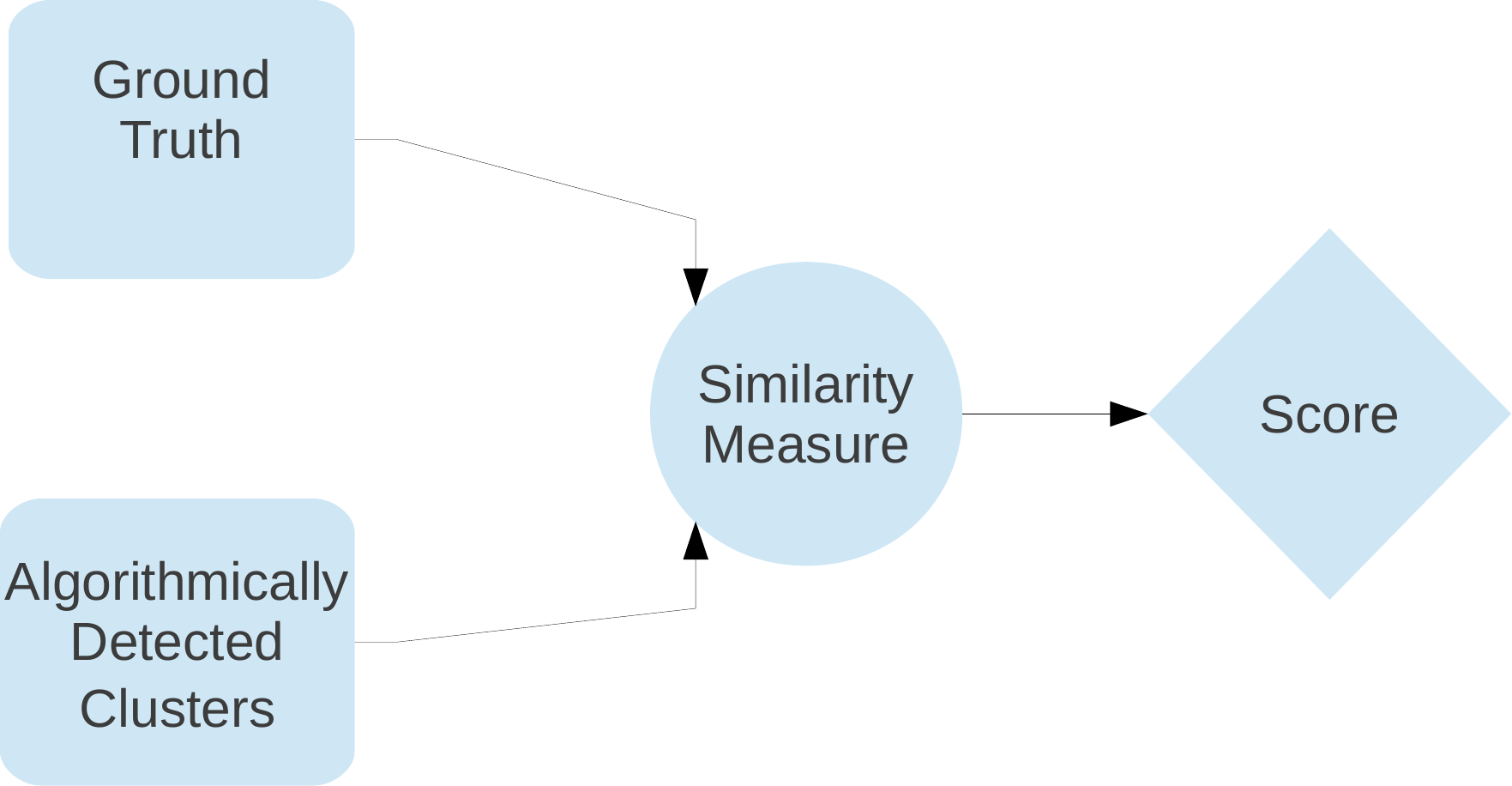}
  \caption{The straightforward way of benchmarking an algorithm using
    a network with a ground-truth set of communities.}
  \label{fig:simple-benchmark}
\end{figure}

Due to these deficiencies, here we focus on the possibility of using
real-world benchmark graphs on digitally extracted data.
Such a benchmark is typically carried out as in \cref{fig:simple-benchmark}.
In the last section, we mentioned that many of the smaller,
hand-curated datasets include a natural ground truth set of
communities which can be used for real-world benchmarks. It is
possible to construct this ground-truth because a researcher (or team
of researchers) has first-hand experience studying a relatively
small social system. For digitally-extracted datasets, no such experts
exist, partly because the process of gathering the dataset (which
typically involves writing a web crawler or log-file parser) does not
involve first-hand contact with a social system, and partly because
the dataset is simply too large for any individual to have detailed
knowledge of its structure.

Recent attempts to create a ground-truth set of communities for a
large, digitally extracted social network typically employ relevant
meta-data---perhaps the most significant work in this
direction is in two recent papers from
Yang and Leskovec \citep{Yang2012a, Yang2012b}, who examine
the social networks of LiveJournal, Orkut, Friendster, and Ning. Each
of these social network datasets includes groups of users
explicitly created by users. Another common type of social
network data that includes an explicit ground-truth set of communities
is co-authorship networks, where attendance at conferences is used to
define the ground truth set of communities.

\subsection{Problem: so-called ground truths are incomplete}
We now come to this section's central question: while one can
certainly choose to define a ground truth set of communities by using
this meta-data, is this a sensible thing to do? In other words: should
we punish a community finding algorithm if it's output does not match
such a ground truth? We argue that the ground truth set of communities
which can be extracted from such meta-data is likely to be woefully
incomplete. 

We substantiate this objection with a concrete example based on the
Facebook100 dataset, introduced by Traud et al. \citep{Traud2011}. We choose our
example from this dataset because because it is in many ways an ideal
dataset for a community detection benchmark: it comes from Facebook,
which at the time the data was collected in 2005, was extremely
popular among college students, and thus the data provides thorough
coverage of the acquaintanceship among students. Furthermore, as the
data was provided directly by Facebook, the dataset is not based on a
sample, but rather includes \emph{all} Facebook friendships. Finally,
it includes meta-data based on the profile page of each user that
indicates dorm membership, gender, graduation year, and academic
major.

As our particular example, we will examine the University of Chicago
sub-network, which includes 6,591 nodes and 208,103 undirected
edges. This sub-network is chosen because one of the authors has
first-hand experience of the social life there as an undergraduate and
is even included in the dataset. Furthermore, the residential
``houses'' are known to be of utmost importance to the social life at
the University of Chicago: upon entering the university, every student
is required to spend at least one year living in the dormitory system,
and the friendships formed in this stage of college---for many
students, the first time living away from home---often endure for
years. According to the University of Chicago website, ``each house
represents a tight-knit community of students, resident faculty
masters, and residential staff, who live, relax, study, dine together
at House Tables, engage, socialize, and learn from each other'' \citep{Housing2}.
\footnote{The meta-data does not explicitly indicate that the ``dorm''
  meta-data represents the housing system at the University of
  Chicago, but this is strongly suggested by the data itself, as the
  number of distinct values corresponds quite closely to the number of
  houses and fraternity houses. At the time of data collection, the
  housing system consisted of 10 residence halls (physical buildings)
  which were further subdivided into 37 houses, which typically
  represent a physically adjacent wing of a residence hall. House
  sizes range from 100 to 37 students, with an average size of
  70 \citep{Housing}. Furthermore, there were somewhere between ten
  and twenty fraternity houses. Thus some of the dorm meta-data may
  indicate fraternity membership.} 

\begin{figure}[!h]
  \centering
    \includegraphics[width=0.75\linewidth]{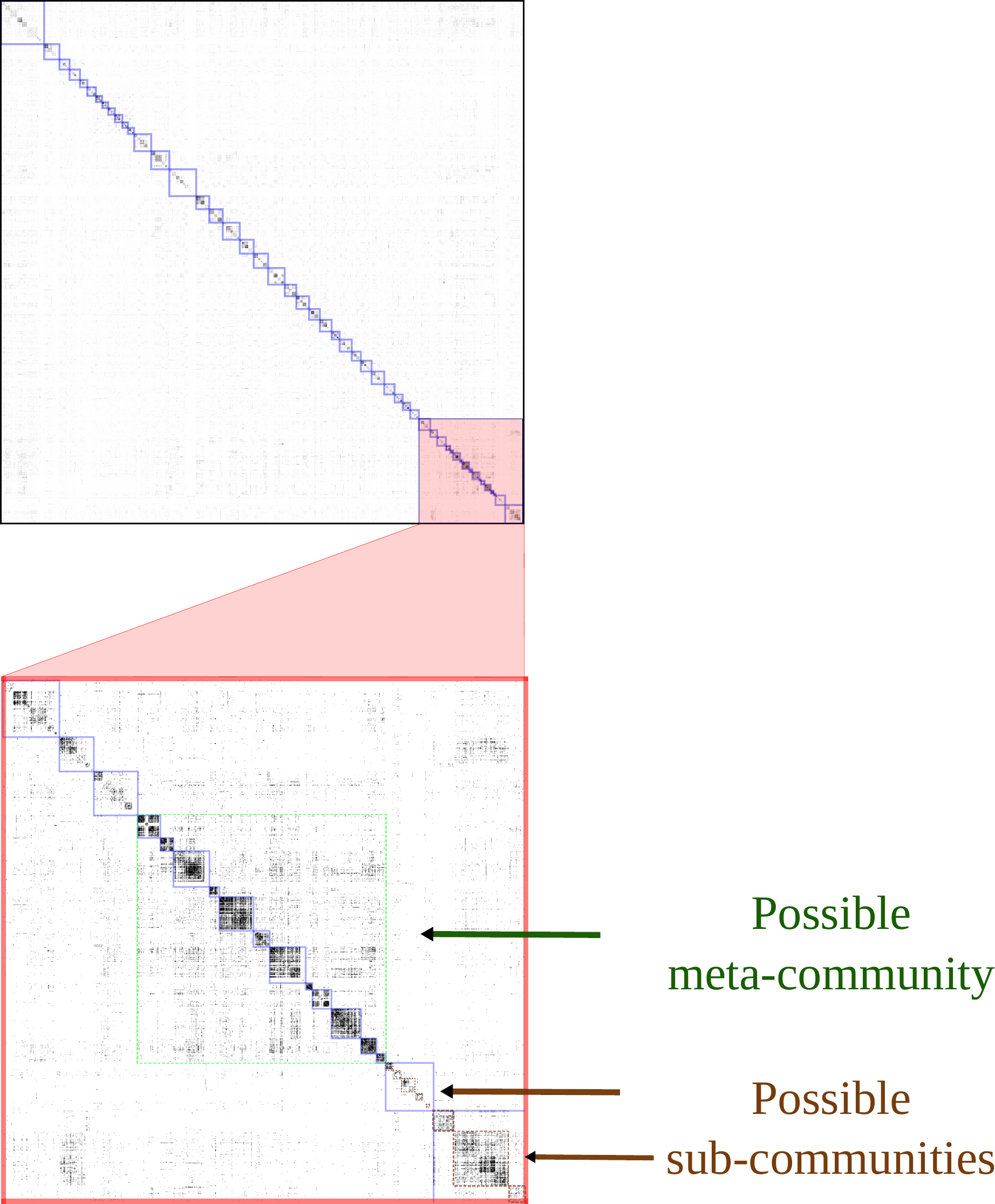}
    \caption{Upper pane: the adjacency matrix of the University of
      Chicago, with house membership highlighted (zoomable
      online). Lower pane: a zoomed in region suggests that
      sub-communities exist within houses, and that macro-structure
      exists between houses.}
    \label{fig:adjacency-matrix}
\end{figure}

The adjacency matrix of this network is displayed in the top pane of
\cref{fig:adjacency-matrix}. The nodes have been ordered so that all
nodes belonging to the same house are adjacent to each other. We can
confirm that the house meta-data is indeed relevant for friendships by
noting when nodes are arranged in this order, dense blocks form
on the main diagonal.

However, to return to the central question of this section: should we
use these dorms as the ground truth set of communities in a benchmark?
In other words: does the grouping of nodes by house correspond to the
grouping of network communities that exist in the network?  The
ordering of nodes in \cref{fig:adjacency-matrix} not only blocks nodes
by house, but also attempts to highlight any sub or
meta-structure.\footnote{To highlight sub-structure within each house,
  we performed the following steps: we first extracted the sub-graph
  induced by each house (i.e., the edges contained in each blue-square
  in \cref{fig:adjacency-matrix}) and then ran a community detection
  method on that sub-graph (we used the Louvain method of modularity
  maximization \citep{Blondel2008}, but many non-overlapping community
  detection algorithms would also have been appropriate). We then
  arranged the nodes within each house block-wise by the network
  communities found in that house's sub-graph. To highlight
  meta-structure between dorms, we created a meta-graph by turning
  each house into one node in the meta graph, and weighted the edges
  between nodes in the meta-graph by the total number of edges which
  connected houses in the original graph. We then found communities on
  the meta-graph using the same method. Finally, we ordered the houses
  in \cref{fig:adjacency-matrix} block-wise by the sets of houses that
  were identified as belonging to the same meta-community.} We can
observe that within many of the houses there is clearly defined
sub-structure. A few examples of this sub-structure are highlighted in
the brown squares in the
bottom pane of \cref{fig:adjacency-matrix}. We
can also observe that several of the houses together form one
meta-community. An example of this meta-community is highlighted in
green in the same image. The twelve houses that belong to this
meta-community could represent houses that are physically
co-located in the same residence hall.

We do not claim to have found the correct ground truth set of
communities in \cref{fig:adjacency-matrix}; there may be alternative
orderings of the adjacency matrix that highlight additional community
structure. The point we want to make is that if we were to simply use
the houses as ground truth communities in a benchmarking setting, then
we would be using, at best, and \emph{incomplete} ground truth,
because it contains neither the sub-structure nor meta-structure that
is clearly visible in the adjacency matrix. We also stress that the
set of houses alone is not even approximately complete, because while
there are only around fifty blue squares, there are several times as
many sub-communities visible. Thus, if a community finding algorithm
were to find all of the communities that are visible in the adjacency
matrix, and then evaluated by measuring how similar those found
communities were to the ground truth of houses, the two sets would not
be very similar according to several common similarity
measures. According such a benchmark, the algorithm would therefore
perform poorly, even though in fact it did a good job of detecting
communities.

Thus, even though on the face of it the housing meta-data appeared to
provide a great ground-truth for a community detection benchmark, it
turns out that it is gravely incomplete. We should therefore not build a
benchmark based on the assumption that there is a one-to-one mapping
between network communities and houses. We also believe that similar
problems apply to other digitally extracted social network datasets.
For example, in the annotated networks used in \citep{Yang2012a,
  Yang2012b}, the ground-truth set of communities is provided
explicitly by users who typically create groups in an online social
network service. There may be many network communities that the users
themselves are not even aware of, or simply do not create explicitly
groups for. Also, users may not be aware of meta-communities, as these
could include thousands of users.


\subsection{Solution: Measure communities' \emph{relation} to meta-data rather than
  \emph{correspondence} with it}
Returning to the University of Chicago example, we ask:
can we create any reasonable benchmark based on the house meta-data?
Certainly the houses are related to the community structure, but we
are unsure of the nature of this relationship. This relationship could
be simple, for example, if a node is a member of a given network
community, then it is more likely to be a member of some house. Or
this relationship could be more complex, for example, if a node
is a member of a certain set of network communities and at the same
time not a member of some other set of network communities, then there is a
high probability of belonging to a certain house. Due to these
possibilities, while we might know that certain
meta-data (such as the housing-meta data) is closely related to
community structure, we often do not know the exact nature of this relationship.

\begin{figure}[h]
  \centering
  \includegraphics[width=0.65\textwidth]{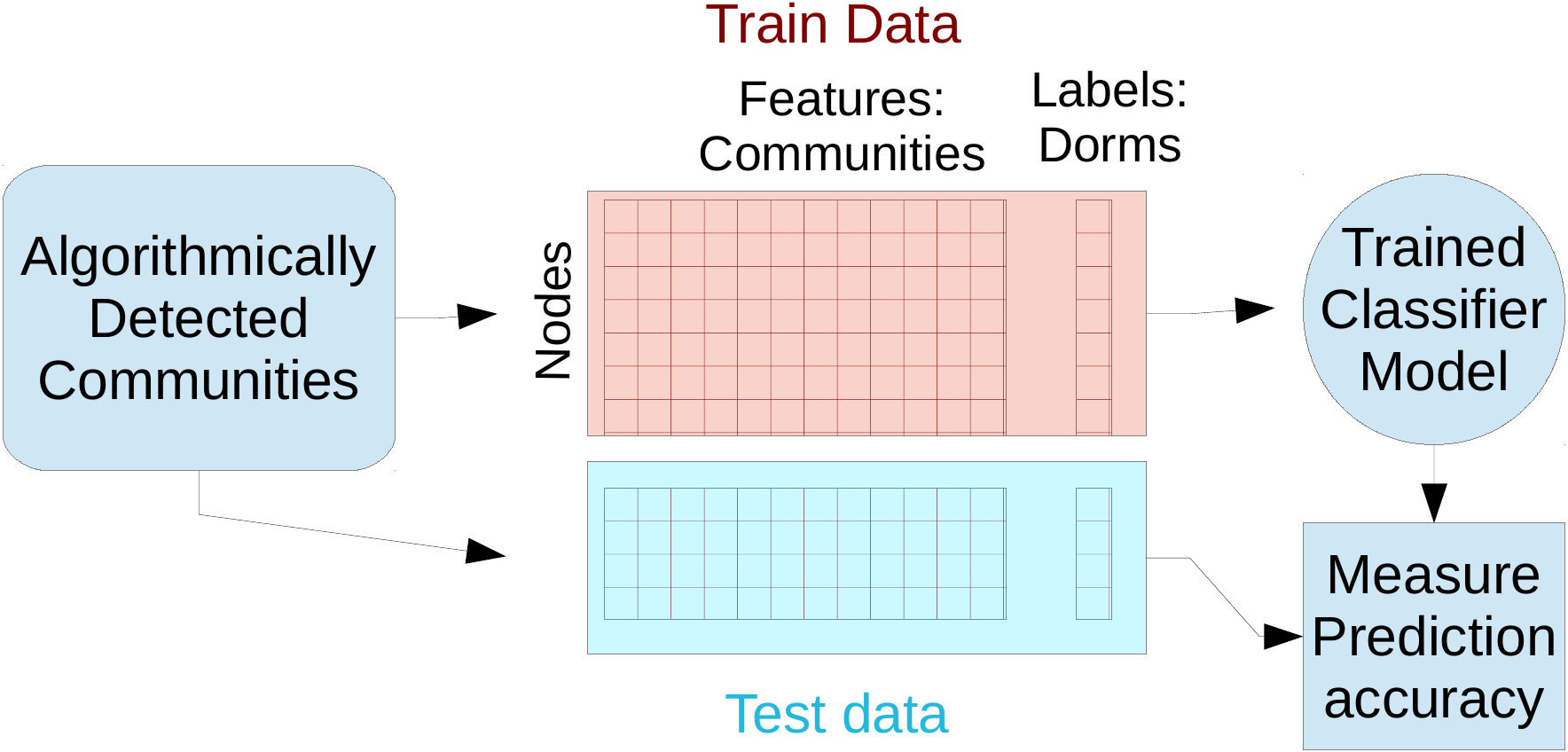}
  \caption{Procedure for a community-detection benchmark based on
    attribute inference.}
  \label{fig:inference-benchmark}
\end{figure}

Thus, we would like to create a benchmarking scheme which allows for a
flexible relationship between network communities and meta-data that
we believe is closely related to community structure. In such
a scheme, we do not want to assume we have knowledge of a complete
ground-truth set of communities.

We observe that the objective of machine learning is to come up with
models that flexibly capture the relationship between a set of
features and some target attribute. In fact, machine learning models
are designed to be used precisely in situations where the relationship
between the features and the target attribute is complex and unknown.
Building on this observation, we propose that one valid way to
incorporate meta-data into a benchmarking scheme is to treat the
meta-data as a node attribute whose value can be more easily inferred
with a good set of network communities. In a way, we shift from a
ground-truth based benchmark to a task-oriented benchmark, where the
task is to infer missing meta data.

The benchmarking procedure that we propose is illustrated in
\cref{fig:inference-benchmark}.  The first step is to detect
communities with the algorithm which is being benchmarked. These
detected communities are used to build a community assignment matrix,
where each row corresponds to a node, and each column corresponds to a
detected community. The values in this matrix are either one or
zero---a value of one at position $(i, j)$ indicates that node $i$ is
a member of community $j$. The meta-data is used to label each node's
class. With the community-assignment matrix as the feature matrix, and
the meta-data as the labels, we can then train a machine learning
model in the usual manner, and use 10-fold cross validation to measure
the accuracy of the model.

The accuracy indicates how well the communities as features allow the
meta-data to be inferred---the higher the accuracy, the better the
communities. This makes sense if we assume that the community
structure is closely related to the meta-data. However, there are some
important conceptual objections one could raise against this
benchmarking scheme. First, all sorts of network features might be
useful for inferring missing meta-data, so the algorithm that performs
best in such a benchmark may not even be detecting network communities
at all. Second, if a community detection algorithm produces several
irrelevant communities, then many machine learning models are clever
enough to simply ignore these.  Thus, a community detection algorithm
will not necessarily be punished for producing a large set of bogus
communities. In fact, if one had infinite computing time and an ideal
machine learning model, then a trivial community detection algorithm
would always obtain the top score: the algorithm that produces all
possible communities. Given all these communities, the ideal machine
learning model would discover which subset of communities allowed the
most accurate inference of the meta-data.

We point out that any benchmarking scheme which is based on an incomplete
ground-truth will have the same problems, because given a detected
community that does not correspond to anything in the incomplete
ground truth, one cannot say whether the community is invalid or
whether it is valid but not included in the ground truth.

Furthermore, while these objections are valid and need to be kept in
mind when interpreting results, we believe they can be addressed. One
can ensure that the algorithm in question does in fact detect network
communities and that it does not detect bogus communities by running
it on synthetic synthetic benchmark networks with clear community
structure in which the ground-truth is complete. In this context, the
synthetic benchmark networks are not used to determine which community
detection algorithm is the best, but rather as a sort of sanity check
to make sure that the algorithm in question is in fact a community
detection algorithm and does not detect a large number of bogus
communities. Thus, the synthetic network used should have clear
community structure that is relatively easy to detect.

One might argue that since the goal here is simply to measure how
related the community assignment matrix is to some meta-data, a
simpler approach would be to use some measure of mutual information
between the community assignment matrix and the meta-data. If the
community structure were a non-overlapping partition, then this task
would be straightforward---one could use the normalized mutual
information as defined in \citep{Danon2005}. Because nodes can belong
to several communities, this measurement becomes more difficult. While
the mutual information of groupings (as opposed to partitions) has
been defined (see \citep{McDaid2011} for an overview), these
definitions do not take interactions between community membership into
account. Because we believe that in social systems the relation
between community membership and some target attribute may be more
complex than allowed for by these measure of mutual information, we
choose instead to use machine learning models to measure this
relation, as these strive to form more flexible hypotheses about how a
feature space is related to some target attribute.

We began this section by defining three types of benchmarks used to
measure the performance of community detection algorithms, and then
explained why one would ideally use real-world networks in which the
complete ground truth was known. We then argued that the meta-data
associated with digitally-extracted networks is unfortunately likely
to be incomplete, and therefore inappropriate for real-world
benchmarks. Thus, while in theory we would like to use real-world data
with complete ground truths, in practice no such datasets
exist. Finally, we proposed an alternative benchmarking scheme which
includes both a task-oriented component and a ``sanity check''
component based on synthetic data. In the next section, we carry out
the proposed task-oriented benchmark in order to provide a concrete
example of what we have in mind, and to demonstrate that such a
benchmarking scheme can reveal insight into the behavior into the
community detection algorithms.


\section{Illustrative example: a benchmark based on Facebook data}
\label{sec:results}
The primary purpose of this section is to flesh out the
benchmarking procedure outlined in
\cref{fig:inference-benchmark}. Along the way we also uncover
problematic behavior exhibited by a few community detection
algorithms; for example, the Louvain and InfoMap methods seem to
detect community structure only at a very coarse level, even if
hierarchical versions of these algorithms are used.

\xhdr{Data and experiment design.}
First, a word on the data. In the last section, we illustrated an
example with a Facebook network representing acquaintanceships at the
University of Chicago. As we mentioned above, this network came from a
larger dataset, the Facebook100 dataset from Traud et al. \citep{Traud2011}, which
includes Facebook data on 100 collegiate networks; these 100 networks
are the data used here. These networks range in size from 769 nodes
and 17k edges to 36k nodes and 1.6m edges. The data has all of the
desirable characteristics described above, for example, it comes
directly from Facebook and is not sampled. Furthermore, the dataset
includes node attribute information on five attributes: gender, year
of graduation, dormitory (as used in the last section), academic
major, and high school. Note that no edges exist between
members of different networks, and that each network is treated
independently of the other networks.

Traud et al. found that two
of these attributes had a close relationship with community structure:
year of graduation and dorm.  Our benchmark will therefore contain two
separate components: one in which communities are used to infer dorm
membership, and the other in which they are used to infer year of
graduation.  For each combination of network, and algorithm, we first
detect communities and use them to build a community assignment
matrix. Next, for each attribute, we use ten-fold cross validation to
measure how well a classifier can infer the attribute based on the
community assignment matrix.\footnote{For some nodes, attribute values
  were missing; i.e., some Facebook users failed to mention on their
  profile which dorm they lived in or which year they graduated. We
  simply leave these nodes out of the cross-validation.} We measure a
classifier's accuracy by simply calculating the percentage of time
that a classifier correctly infers the attribute.

We reiterate that the assumption underlying benchmarks is that
each of these attributes is closely related to community structure,
and so as the community assignment matrix tends to improve and
more closely resemble the unknown ground truth, this matrix will allow
a machine learning classifier to more accurate inference of missing
attributes.

\xhdr{Classifier.}
Some of the community detection methods benchmarked here detect
thousands of communities on these networks, so it is essential that
the classifier used performs well in situations with thousands of
features, otherwise our benchmark may be biased against methods which
detect many communities. After experimenting with several classifiers
and feature selection schemes, we found that an ensemble method called
stochastic gradient boosting both performed best and was least
sensitive to large numbers of communities. In particular, we use the
implementation provided in the Python package scikit-learn
\citep{scikit-learn}, with the learning rate set to 0.005 and the
number of trees set to 1000.\footnote{There were two more parameter
  values to set: we required at least five examples for a split in a
  decision tree (i.e., the max\_samples\_subsplit parameter$=5$) and
  set the subsampling rate to $0.4$. We arrived at these values
  through experimentation, choosing those values which maximized the
  performance of the classifier.}

\xhdr{Community detection methods tested.}
We benchmark four community detection algorithms: the Louvain method
of modularity maximization \citep{Blondel2008}, the InfoMap method of map equation
maximization \citep{Rosvall2011}, the Link Community method (LC) \citep{Ahn2010}, and the Greedy Clique
Expansion algorithm (GCE) \citep{Lee2010}. We choose the Louvain method and InfoMap
because they are perhaps the two currently most popular methods of
community detection.  We include the Link Community method and GCE
because they both claim to handle the case of overlapping communities
particularly well, and we have reason to believe that in the Facebook
data most nodes could belong to multiple communities.

We used the author's implementation of the Louvain
method\footnote{\url{https://sites.google.com/site/findcommunities/}},
which allows for both flat and hierarchical partitions, both of which
will be considered below.  We also used the author's implementation of
the InfoMap
algorithm\footnote{\url{http://www.tp.umu.se/~rosvall/code.html}}
presented in \citep{Rosvall2011}, which is designed to detect
hierarchical community structure.  Likewise, we used the author's C++
implementation of LC, which can detect either a flat or hierarchical
clustering, both of which will be tested
below.\footnote{\url{https://github.com/bagrow/linkcomm}} Because LC
often found vast numbers of extremely small communities, we removed
all communities containing fewer than four nodes or three edges. For GCE, we
used the author's
implementation,\footnote{\url{https://sites.google.com/site/greedycliqueexpansion/}}
and set the value of the resolution parameter $\alpha$ to 1.5, as this
value was recommended for the Facebook data in previous work. We
should note that because GCE has a resolution that has been tuned to
this type of data, it has an unfair advantage over the other
algorithms; however, in the latter part of this section we also try to
optimally tune the resolution parameters of the other methods.

\begin{table}
  \begin{center}
  \def\infohierdorm{data/infohier-dorm-scores.txt}
\def\infohieryear{data/infohier-year-scores.txt}
\def\infohierUChicagoSmallest{data/infohier-UChicagoSmallest.txt}
\def\louvaintwoadorm{data/louvain02a-dorm-scores.txt}
\def\louvaintwoayear{data/louvain02a-year-scores.txt}
\def\louvaintwoaUChicagoSmallest{data/louvain02a-UChicagoSmallest.txt}
\def\louvaintendorm{data/louvain10-dorm-scores.txt}
\def\louvaintenyear{data/louvain10-year-scores.txt}
\def\louvaintenUChicagoSmallest{data/louvain10-UChicagoSmallest.txt}
\def\linkClusterCombineddorm{data/linkClusterCombined-dorm-scores.txt}
\def\linkClusterCombinedyear{data/linkClusterCombined-year-scores.txt}
\def\linkClusterCombinedUChicagoSmallest{data/linkClusterCombined-UChicagoSmallest.txt}
\def\timescaleCombinedEfivehundreddorm{data/timescaleCombinedE500-dorm-scores.txt}
\def\timescaleCombinedEfivehundredyear{data/timescaleCombinedE500-year-scores.txt}
\def\timescaleCombinedEfivehundredUChicagoSmallest{data/timescaleCombinedE500-UChicagoSmallest.txt}
\def\gceCombinedEfivehundreddorm{data/gceCombinedE500-dorm-scores.txt}
\def\gceCombinedEfivehundredyear{data/gceCombinedE500-year-scores.txt}
\def\gceCombinedEfivehundredUChicagoSmallest{data/gceCombinedE500-UChicagoSmallest.txt}
\def\louvaintenadorm{data/louvain10a-dorm-scores.txt}
\def\louvaintenayear{data/louvain10a-year-scores.txt}
\def\louvaintenaUChicagoSmallest{data/louvain10a-UChicagoSmallest.txt}
\def\linkclusterdorm{data/linkcluster-dorm-scores.txt}
\def\linkclusteryear{data/linkcluster-year-scores.txt}
\def\linkclusterUChicagoSmallest{data/linkcluster-UChicagoSmallest.txt}
\def\gcefifteendorm{data/gce15-dorm-scores.txt}
\def\gcefifteenyear{data/gce15-year-scores.txt}
\def\gcefifteenUChicagoSmallest{data/gce15-UChicagoSmallest.txt}
\def\louvainfiveadorm{data/louvain05a-dorm-scores.txt}
\def\louvainfiveayear{data/louvain05a-year-scores.txt}
\def\louvainfiveaUChicagoSmallest{data/louvain05a-UChicagoSmallest.txt}
\begin{tabular}{lcccccc}
\toprule
 & \MC{Dorm Accuracies} & \MC{Year Accuracies} & \MC{UChicago Stats}\\
\cmidrule(lr){2-3} \cmidrule(lr){4-5} \cmidrule(lr){6-7}
\textbf{Method} & \textbf{Histogram} & \textbf{Mean} & \textbf{Histogram} & \textbf{Mean} & \textbf{Median Smallest} & \textbf{\# Comms} \\
\midrule
    gce15     & 
    \begin{tikzpicture}[xscale=0.01, yscale=0.1]
        \draw[ultra thin, black!50] (1,0)--(100,0);
        \begin{scope}[ycomb, yscale=0.30]
            \draw[black, thin] plot[] file {\gcefifteendorm};
        \end{scope}
    \end{tikzpicture}
    & 47.0    & 
    \begin{tikzpicture}[xscale=0.01, yscale=0.1]
        \draw[ultra thin, black!50] (1,0)--(100,0);
        \begin{scope}[ycomb, yscale=0.30]
            \draw[black, thin] plot[] file {\gcefifteenyear};
        \end{scope}
    \end{tikzpicture}
    & 65.0     & 43.0     & 266 \\
    infohier     & 
    \begin{tikzpicture}[xscale=0.01, yscale=0.1]
        \draw[ultra thin, black!50] (1,0)--(100,0);
        \begin{scope}[ycomb, yscale=0.30]
            \draw[black, thin] plot[] file {\infohierdorm};
        \end{scope}
    \end{tikzpicture}
    & 32.7    & 
    \begin{tikzpicture}[xscale=0.01, yscale=0.1]
        \draw[ultra thin, black!50] (1,0)--(100,0);
        \begin{scope}[ycomb, yscale=0.30]
            \draw[black, thin] plot[] file {\infohieryear};
        \end{scope}
    \end{tikzpicture}
    & 57.2     & 171.0     & 114 \\
    louvain10     & 
    \begin{tikzpicture}[xscale=0.01, yscale=0.1]
        \draw[ultra thin, black!50] (1,0)--(100,0);
        \begin{scope}[ycomb, yscale=0.30]
            \draw[black, thin] plot[] file {\louvaintendorm};
        \end{scope}
    \end{tikzpicture}
    & 25.4    & 
    \begin{tikzpicture}[xscale=0.01, yscale=0.1]
        \draw[ultra thin, black!50] (1,0)--(100,0);
        \begin{scope}[ycomb, yscale=0.30]
            \draw[black, thin] plot[] file {\louvaintenyear};
        \end{scope}
    \end{tikzpicture}
    & 60.0     & 1016.0     & 27 \\
    linkcluster     & 
    \begin{tikzpicture}[xscale=0.01, yscale=0.1]
        \draw[ultra thin, black!50] (1,0)--(100,0);
        \begin{scope}[ycomb, yscale=0.30]
            \draw[black, thin] plot[] file {\linkclusterdorm};
        \end{scope}
    \end{tikzpicture}
    & 27.9    & 
    \begin{tikzpicture}[xscale=0.01, yscale=0.1]
        \draw[ultra thin, black!50] (1,0)--(100,0);
        \begin{scope}[ycomb, yscale=0.30]
            \draw[black, thin] plot[] file {\linkclusteryear};
        \end{scope}
    \end{tikzpicture}
    & 38.5     & 4.0     & 10 \\
\bottomrule
\end{tabular}

  \caption{Performance of four community detection algorithms on
    inferring dorm and year of graduation. The ``UChicago Stats''
    columns indicate some relevant statistics about communities found
    on the University of Chicago's Facebook network, whose adjacency
    matrix is displayed in \cref{fig:adjacency-matrix}.}
  \label{table:basic}
\end{center}
\end{table}

\xhdr{Results.}  Results are presented in \cref{table:basic}.  Note
that training the classifier is computationally expensive, and for
this reason we run our benchmark on only the forty smallest
universities. Furthermore, rather than carrying out the evaluation of
accuracy on all ten folds of the cross-validation scheme, we use only
three. \footnote{Even so, our benchmark scheme took months to carry
  out on a machine with 32 cores---it required us to train 2400
  classifiers (ten community detection methods, three folds, forty
  universities, two attributes). Some of these classifiers were slow
  to train because they needed to be trained on more than ten thousand
  features (a variation of the link-clustering method labeled below as
  \emph{linkClusterCombined} found by far the most communities and
  therefore the classifiers trained on these communities took up most
  of the CPU time).}  Thus, for each combination of community
detection method and attribute, we detected communities on the forty
smallest networks, then trained classifiers to infer the attribute on
three folds, yielding a total of 120 classifiers. The ``mean
accuracy'' columns presented in
\cref{table:basic,table:multi,table:multiCombined} therefore indicate
the average accuracy obtained by these 120 classifiers. We also show
the distribution of these 120 accuracies with a histogram.

Turning to the results, we see that GCE has the best performance on
inferring values for both the dorm and year attributes; in particular,
it performs substantially better than the other methods on the dorm inference
task.  We observe that the Louvain and InfoMap methods detected a
smaller number of larger communities, whereas GCE tended to find more
and smaller communities.  We believe that the Louvain and InfoMap
methods performed poorly because they missed the fine-grained
community structure.

We can demonstrate this point by returning to the example of the
University of Chicago that was presented in
\cref{fig:adjacency-matrix}. The ``UChicago Stats'' section of the
table indicates the number of communities found on the University of
Chicago's Facebook network---this is helpful because we have already
seen from the adjacency matrix in \cref{fig:adjacency-matrix} that
this network contains at least several dozen (perhaps hundreds) of
network communities, and that nodes appear to belong to both small and
large communities. To calculate the ``median smallest'' column, for
each node we first recorded the size of the smallest community it
belonged to, and then we took the median of this distribution.  This
column indicates that the Louvain and InfoMap methods failed to place
most nodes into any small communities, whereas GCE did typically place
nodes into smaller communities. We would expect these fine-grained
communities to be useful for inferring dorm attribute, because the
partition formed by dorm membership includes more and smaller groups
than the one formed by the year attribute.

The LC method behaved strangely on the UChicago network and several of
the other networks: the vast majority of the communities found were
signleton communities (containing one edge and two nodes), and only
ten of the detected communities contained four or more nodes. It
appears that the cut made in the dendrogram produced by this network
was made at too low of a level---we will try to remedy this problem
below by making cuts at several levels and combining the communities
from each cut. First, we will see if we can improve the performance of
the Louvain method by tuning it to detect finer-scale structure.

\begin{table}
  \centering
  \def\infohierdorm{data/infohier-dorm-scores.txt}
\def\infohieryear{data/infohier-year-scores.txt}
\def\infohierUChicagoSmallest{data/infohier-UChicagoSmallest.txt}
\def\louvaintwoadorm{data/louvain02a-dorm-scores.txt}
\def\louvaintwoayear{data/louvain02a-year-scores.txt}
\def\louvaintwoaUChicagoSmallest{data/louvain02a-UChicagoSmallest.txt}
\def\louvaintendorm{data/louvain10-dorm-scores.txt}
\def\louvaintenyear{data/louvain10-year-scores.txt}
\def\louvaintenUChicagoSmallest{data/louvain10-UChicagoSmallest.txt}
\def\linkClusterCombineddorm{data/linkClusterCombined-dorm-scores.txt}
\def\linkClusterCombinedyear{data/linkClusterCombined-year-scores.txt}
\def\linkClusterCombinedUChicagoSmallest{data/linkClusterCombined-UChicagoSmallest.txt}
\def\timescaleCombinedEfivehundreddorm{data/timescaleCombinedE500-dorm-scores.txt}
\def\timescaleCombinedEfivehundredyear{data/timescaleCombinedE500-year-scores.txt}
\def\timescaleCombinedEfivehundredUChicagoSmallest{data/timescaleCombinedE500-UChicagoSmallest.txt}
\def\gceCombinedEfivehundreddorm{data/gceCombinedE500-dorm-scores.txt}
\def\gceCombinedEfivehundredyear{data/gceCombinedE500-year-scores.txt}
\def\gceCombinedEfivehundredUChicagoSmallest{data/gceCombinedE500-UChicagoSmallest.txt}
\def\louvaintenadorm{data/louvain10a-dorm-scores.txt}
\def\louvaintenayear{data/louvain10a-year-scores.txt}
\def\louvaintenaUChicagoSmallest{data/louvain10a-UChicagoSmallest.txt}
\def\linkclusterdorm{data/linkcluster-dorm-scores.txt}
\def\linkclusteryear{data/linkcluster-year-scores.txt}
\def\linkclusterUChicagoSmallest{data/linkcluster-UChicagoSmallest.txt}
\def\gcefifteendorm{data/gce15-dorm-scores.txt}
\def\gcefifteenyear{data/gce15-year-scores.txt}
\def\gcefifteenUChicagoSmallest{data/gce15-UChicagoSmallest.txt}
\def\louvainfiveadorm{data/louvain05a-dorm-scores.txt}
\def\louvainfiveayear{data/louvain05a-year-scores.txt}
\def\louvainfiveaUChicagoSmallest{data/louvain05a-UChicagoSmallest.txt}
\begin{tabular}{cccccccc}
\toprule
\MC{Louvain Parameters} & \MC{Dorm Accuracies} & \MC{Year Accuracies} & \MC{UChicago Stats}\\
\cmidrule(lr){1-2} \cmidrule(lr){3-4} \cmidrule(lr){5-6} \cmidrule(lr){7-8}
\textbf{Markov Time} & \textbf{Multi-level} & \textbf{Histogram} & \textbf{Mean} & \textbf{Histogram} & \textbf{Mean} & \textbf{Median Smallest} & \textbf{\# Comms} \\
\midrule
    1.0     &      & 
    \begin{tikzpicture}[xscale=0.01, yscale=0.1]
        \draw[ultra thin, black!50] (1,0)--(100,0);
        \begin{scope}[ycomb, yscale=0.30]
            \draw[black, thin] plot[] file {\louvaintendorm};
        \end{scope}
    \end{tikzpicture}
    & 25.4    & 
    \begin{tikzpicture}[xscale=0.01, yscale=0.1]
        \draw[ultra thin, black!50] (1,0)--(100,0);
        \begin{scope}[ycomb, yscale=0.30]
            \draw[black, thin] plot[] file {\louvaintenyear};
        \end{scope}
    \end{tikzpicture}
    & 60.0     & 1016.0     & 27 \\
    1.0     & \checkmark    & 
    \begin{tikzpicture}[xscale=0.01, yscale=0.1]
        \draw[ultra thin, black!50] (1,0)--(100,0);
        \begin{scope}[ycomb, yscale=0.30]
            \draw[black, thin] plot[] file {\louvaintenadorm};
        \end{scope}
    \end{tikzpicture}
    & 27.2    & 
    \begin{tikzpicture}[xscale=0.01, yscale=0.1]
        \draw[ultra thin, black!50] (1,0)--(100,0);
        \begin{scope}[ycomb, yscale=0.30]
            \draw[black, thin] plot[] file {\louvaintenayear};
        \end{scope}
    \end{tikzpicture}
    & 61.1     & 923.0     & 97 \\
    0.5     & \checkmark    & 
    \begin{tikzpicture}[xscale=0.01, yscale=0.1]
        \draw[ultra thin, black!50] (1,0)--(100,0);
        \begin{scope}[ycomb, yscale=0.30]
            \draw[black, thin] plot[] file {\louvainfiveadorm};
        \end{scope}
    \end{tikzpicture}
    & 33.5    & 
    \begin{tikzpicture}[xscale=0.01, yscale=0.1]
        \draw[ultra thin, black!50] (1,0)--(100,0);
        \begin{scope}[ycomb, yscale=0.30]
            \draw[black, thin] plot[] file {\louvainfiveayear};
        \end{scope}
    \end{tikzpicture}
    & 62.2     & 231.0     & 151 \\
    0.2     & \checkmark    & 
    \begin{tikzpicture}[xscale=0.01, yscale=0.1]
        \draw[ultra thin, black!50] (1,0)--(100,0);
        \begin{scope}[ycomb, yscale=0.30]
            \draw[black, thin] plot[] file {\louvaintwoadorm};
        \end{scope}
    \end{tikzpicture}
    & 42.7    & 
    \begin{tikzpicture}[xscale=0.01, yscale=0.1]
        \draw[ultra thin, black!50] (1,0)--(100,0);
        \begin{scope}[ycomb, yscale=0.30]
            \draw[black, thin] plot[] file {\louvaintwoayear};
        \end{scope}
    \end{tikzpicture}
    & 63.5     & 93.0     & 254 \\
\bottomrule
\end{tabular}

  \caption{Performance of variations of the Louvain method of
    modularity maximization---the left-most column indicates the value
    of ``Markov time'' used. Markov time is a resolution parameter;
    when set to 1.0, the original definition of modularity is
    recovered. The second column indicates whether a flat cut was used
    on the dendrogram, or all inner nodes of the dendrogram were used
    as communities.}
  \label{table:multi}
\end{table}
\xhdr{The Louvain method and multi-scale community detection.}  To
further investigate whether the Louvain method's poor performance is
due to missing communities at the smallest scale, we perform
additional experiments. The scores for the Louvain method presented in
\cref{table:basic} are based on the optimal flat
partitioning. However, as the Louvain method is based on an
agglomerative, hierarchical clustering, one can also include
communities from all levels of the dendrogram, not just the flat cut
which optimizes modularity. Blondel et al., the authors of the Louvain
method \citep{Blondel2008}, claim that the method ``unfolds a complete
hierarchical community structure for the network,'' which suggests
that the algorithm should detect community structures on all
scales. In the second row of \cref{table:multi}, we present the
results when communities detected at all levels are used. We note that
the accuracy increases slightly, and that the number of communities
found increases---for example, on the University of Chicago network,
the number of communities increased from 27 to 97.

The findings of \citep{Delvenne2010} indicate that to find community
structure at all resolutions, the very definition of modularity should
be parameterized with a parameter called ``Markov time.''  We test
this claim by checking whether such a parameterized version of
modularity can yield fine-grained communities that improve
accuracy. We use the implementation by Renault Lambiotte, which is
fortunately based on the very same implementation of the Louvain
method and so allows for direct comparison.\footnote{Available at
  \url{http://www.lambiotte.be/codes.html}} We set the resolution
parameter to 0.5 and 0.2, which are values that should detect
community structure on a smaller scale than the unparameterized
version of modularity used above, which implicitly sets this value
to 1.0. For each of these values, we extract all communities from the
dendrogram, as described in the last paragraph. We observe that when
the Markov time is decreased, the number of communities detected
increases and the accuracy increases significantly for the dorm
attribute. As the dorm attribute is more closely associated with
finer-scale community structure, this indicates that the resolution
parameter does indeed help to find community structure on a smaller
scale.

This finding suggests that when modularity maximization techniques are
used, then in order to find community structure at smaller scales, it
is not enough simply to use a hierarchical clustering technique and
make cuts at all levels in the resulting dendrogram. Rather, the very
definition of modularity itself must be parameterized with a
resolution parameter. While there is much theoretical literature on
``resolution limit'' inherent in modularity, here we find strong
empirical evidence of this limit.

Our findings here also place the results of Traud et al. \citep{Traud2011} into
doubt. They analyzed the community structure in the
Facebook100 network using a modularity maximization technique, but
paid no consideration to the resolution limit. Traud et al. found
that in larger universities, year of graduation was more relevant for
community structure than dormotory assignment. Our results indicate
that this finding is likely not inherent in the data, but rather due
to a limitation of modularity maxmimization techniques: in larger
networks, a na\"ive application of these techniques does not detect
finer-grained communities.

The importance of a resolution parameter for modularity raises the
question of whether InfoMap could also perform better if its objective
function, the Map Equation, were parameterized with a resolution
parameter. As mentioned above, while the implementation of InfoMap
that we used is designed to detect community structure at all relevant
resolutions, tended to detect only larger communities. Along the lines
of the parameterized definition of modularity discussed above, recent
work in \citep{Schaub2012} has parameterized the Map Equation (which
InfoMap optimizes) with a resolution parameter by modifying the Markov
time used to compute the stationary distribution of the random
walk. We tried to use their implementation, but encountered unexpected
behavior and results that were worse than with the unparameterized
InfoMap.\footnote{In particular, when the resolution parameter of the
  parameterized InfoMap is set to 1.0, then the original InfoMap
  should be recovered, but this was not the case.} We believe that
these results are related to implementation rather than the conceptual
modifications, and therefore do not report these results.

\xhdr{Combining multiple runs to find structure at all scales.}
Leaving InfoMap aside, each of the three other algorithms has a
resolution parameter: for the Louvain method, we have the Markov time,
for the LC method we have the threshold at which to cut the
hierarchical clustering of edges, and GCE has a parameter $\alpha$
which is built into its local objective function. In order to detect
communities at all scales, one could run the algorithm multiple
times using different values for the resolution parameter, and then
combine the results. In our final experiment, we check whether such a
procedure increases the performance on the benchmark. We combine runs
of the Louvain method where the Markov time is set to $t=0.1, t=0.2,
...  , t=1.0$ and ; we combine runs of LC where the threshold for the
cut is set to each integer-valued percentage point between 1 and 100,
and we combine runs of GCE where $\alpha$ is set to 0.8, 1.0, 1.3,
1.5, 1.7, and 2.2.

When combining several runs of a community detection algorithm, the
resulting set of communities can contain a very large set of
near-duplicate communities. For example, in a single run, the LC
method finds over 100,000 communities on some of the Facebook100
networks, so when several runs are combined, this number can reach
into the millions. The vast majority of these millions of communities
are near-duplicates of other communities, an undesirable property in
most settings, and one which in the current context makes the training
of the classifier computationally expensive. When we combine several
runs of an algorithm, we therefore remove the near duplicates by
following the procedure outlined in section 2 of \citep{Lee2010},
setting $\epsilon$ to 0.5; this technique basically removes
communities that have a Jaccard similarity of greater than 0.5 with
any communities of equal or lesser size.

\begin{table}
  \centering
  \def\infohierdorm{data/infohier-dorm-scores.txt}
\def\infohieryear{data/infohier-year-scores.txt}
\def\infohierUChicagoSmallest{data/infohier-UChicagoSmallest.txt}
\def\louvaintwoadorm{data/louvain02a-dorm-scores.txt}
\def\louvaintwoayear{data/louvain02a-year-scores.txt}
\def\louvaintwoaUChicagoSmallest{data/louvain02a-UChicagoSmallest.txt}
\def\louvaintendorm{data/louvain10-dorm-scores.txt}
\def\louvaintenyear{data/louvain10-year-scores.txt}
\def\louvaintenUChicagoSmallest{data/louvain10-UChicagoSmallest.txt}
\def\linkClusterCombineddorm{data/linkClusterCombined-dorm-scores.txt}
\def\linkClusterCombinedyear{data/linkClusterCombined-year-scores.txt}
\def\linkClusterCombinedUChicagoSmallest{data/linkClusterCombined-UChicagoSmallest.txt}
\def\timescaleCombinedEfivehundreddorm{data/timescaleCombinedE500-dorm-scores.txt}
\def\timescaleCombinedEfivehundredyear{data/timescaleCombinedE500-year-scores.txt}
\def\timescaleCombinedEfivehundredUChicagoSmallest{data/timescaleCombinedE500-UChicagoSmallest.txt}
\def\gceCombinedEfivehundreddorm{data/gceCombinedE500-dorm-scores.txt}
\def\gceCombinedEfivehundredyear{data/gceCombinedE500-year-scores.txt}
\def\gceCombinedEfivehundredUChicagoSmallest{data/gceCombinedE500-UChicagoSmallest.txt}
\def\louvaintenadorm{data/louvain10a-dorm-scores.txt}
\def\louvaintenayear{data/louvain10a-year-scores.txt}
\def\louvaintenaUChicagoSmallest{data/louvain10a-UChicagoSmallest.txt}
\def\linkclusterdorm{data/linkcluster-dorm-scores.txt}
\def\linkclusteryear{data/linkcluster-year-scores.txt}
\def\linkclusterUChicagoSmallest{data/linkcluster-UChicagoSmallest.txt}
\def\gcefifteendorm{data/gce15-dorm-scores.txt}
\def\gcefifteenyear{data/gce15-year-scores.txt}
\def\gcefifteenUChicagoSmallest{data/gce15-UChicagoSmallest.txt}
\def\louvainfiveadorm{data/louvain05a-dorm-scores.txt}
\def\louvainfiveayear{data/louvain05a-year-scores.txt}
\def\louvainfiveaUChicagoSmallest{data/louvain05a-UChicagoSmallest.txt}
\begin{tabular}{lcccccc}
\toprule
 & \MC{Dorm Accuracies} & \MC{Year Accuracies} & \MC{UChicago Stats}\\
\cmidrule(lr){2-3} \cmidrule(lr){4-5} \cmidrule(lr){6-7}
\textbf{Method} & \textbf{Histogram} & \textbf{Mean} & \textbf{Histogram} & \textbf{Mean} & \textbf{Median Smallest} & \textbf{\# Comms} \\
\midrule
    linkClusterCombined     & 
    \begin{tikzpicture}[xscale=0.01, yscale=0.1]
        \draw[ultra thin, black!50] (1,0)--(100,0);
        \begin{scope}[ycomb, yscale=0.30]
            \draw[black, thin] plot[] file {\linkClusterCombineddorm};
        \end{scope}
    \end{tikzpicture}
    & 44.7    & 
    \begin{tikzpicture}[xscale=0.01, yscale=0.1]
        \draw[ultra thin, black!50] (1,0)--(100,0);
        \begin{scope}[ycomb, yscale=0.30]
            \draw[black, thin] plot[] file {\linkClusterCombinedyear};
        \end{scope}
    \end{tikzpicture}
    & 61.3     & 5.0     & 15731 \\
    gceCombinedE500     & 
    \begin{tikzpicture}[xscale=0.01, yscale=0.1]
        \draw[ultra thin, black!50] (1,0)--(100,0);
        \begin{scope}[ycomb, yscale=0.30]
            \draw[black, thin] plot[] file {\gceCombinedEfivehundreddorm};
        \end{scope}
    \end{tikzpicture}
    & 53.4    & 
    \begin{tikzpicture}[xscale=0.01, yscale=0.1]
        \draw[ultra thin, black!50] (1,0)--(100,0);
        \begin{scope}[ycomb, yscale=0.30]
            \draw[black, thin] plot[] file {\gceCombinedEfivehundredyear};
        \end{scope}
    \end{tikzpicture}
    & 75.3     & 22.0     & 890 \\
    louvainCombined     & 
    \begin{tikzpicture}[xscale=0.01, yscale=0.1]
        \draw[ultra thin, black!50] (1,0)--(100,0);
        \begin{scope}[ycomb, yscale=0.30]
            \draw[black, thin] plot[] file {\timescaleCombinedEfivehundreddorm};
        \end{scope}
    \end{tikzpicture}
    & 50.8    & 
    \begin{tikzpicture}[xscale=0.01, yscale=0.1]
        \draw[ultra thin, black!50] (1,0)--(100,0);
        \begin{scope}[ycomb, yscale=0.30]
            \draw[black, thin] plot[] file {\timescaleCombinedEfivehundredyear};
        \end{scope}
    \end{tikzpicture}
    & 73.6     & 59.0     & 278 \\
\bottomrule
\end{tabular}

  \caption{Results for when each algorithm is run several
    times---with different values for the resolution
    parameter--and the distinct communities from all runs are
    combined. Note the large improvement for the dorm attribute when
    compared to \cref{table:basic}.}
  \label{table:multiCombined}
\end{table}

The results of this final experiment are displayed in
\cref{table:multiCombined}. We see that each method has benefitted
by combining the results of multiple runs at different settings of the
resolution parameter.

We now wrap up this demonstration benchmark with a summary of our
findings. Many community detection techniques strive to be parameter free so
that they can automatically detect communities without requiring a
user to experiment with different parameter
values \citep{Fortunato2010}. While this is a worthy goal, the results
of this section indicate that to achieve good performance, one must
tweak the resolution parameter of every method tested here. If we
compare the results in \cref{table:basic} with those in
\cref{table:multiCombined}, we see that if one simply trusts the
algorithm to automatically set the resolution parameter, then the
method may in practice struggle to find structure at all relevant
levels. For example, the naive application of the Louvain method
produces a set of communities which allow a classifier to infer the
dorm attribute with an accuracy of only 25.6\%, whereas by combining
multiple runs with different values of the Markov time parameter, one
can obtain an accuracy of 50.4\%. This dramatic increase in
performance (as well as our analysis above) indicates that the naive
application of the algorithm failed to detect much of the
finer-grained community structure.


\section{Conclusion}
In \cref{sec:history}, we distinguished between digitally extracted
networks and small, hand-curated networks, and we argued that these
two types of data differ in important ways. We also pointed out that
although the recent wave of community detection methods are supposed
to work on digitally extracted networks, in practice the only real
data they are tested on are small, hand-curated networks.  As a
result, we are unaware of whether these methods work on digitally
extracted networks. This situation is caused in large part by the
lack of digitally extracted networks with acceptable ground-truth
data; indeed, in \cref{sec:annotated} we showed that even in cases
where it appears that one may have a reasonable ground truth set of
communities, this ground truth is likely to be quite incomplete, and
therefore unsuitable for a straightforward benchmark such as the one
depicted in \cref{fig:simple-benchmark}. In that section we also
proposed an alternative benchmarking scheme that is appropriate for
the case where one has only an incomplete ground-truth. Finally, in
\cref{sec:results} we employed this alternative benchmarking
scheme.

The results in \cref{sec:results} demonstrate how the inference-based
benchmarking scheme we proposed in \cref{sec:annotated} can reveal
limitations of community finding algorithms. The benchmarks indicate that
some of the most popular community detection methods struggle to
detect communities at smaller scales. We note that this problem did
not emerge when these methods were benchmarked on small, hand-curated
networks, and this suggests that we cannot assume that just because a
community detection method works well on small, hand-curated networks
like Zachary's Karate Club, it does not mean it will work as well on
digitally-extracted networks.

An unfortunate drawback of the benchmarking approach
presented here is its complexity. The benchmarking workflow involves
components that are not directly related to community detection---such
as classifiers---which add extra parameters whose values must
be set with care. While this complexity is unfortunate, we
feel that the simpler approach of treating meta-data as if it contained
a complete ground truth, as in \citep{Yang2012a, Yang2012b}, is even
more problematic because it may unfairly punish a community algorithm
for detecting a valid network community that does not exist in an
incomplete ground truth.

One benefit of this benchmarking approach is that it indicates a
practical problem for which network communities are useful. We can
confirm that network communities are in fact useful for inferring
missing attribute values.\footnote{The accuracy with which we could
infer missing node attributes should be taken with a grain of salt. In
our scheme, we measured this accuracy by holding out data from nodes
whose labels were known, whereas in practice, one would want to infer
values for nodes with unknown labels. It could be these nodes with
missing labels differ from the nodes with known labels, and as a
result, the accuracy of inferring their labels might differ. Because
our objective was simply to measure how related community structure is
to meta-data---and not to accurately measure how well we can infer
missing data in practice---this limitation is not problematic for the
work presented here, but should be borne in mind by those who are
interested in attribute inference for its own sake.}

While it is not the primary concern of this paper and so we will not
go into great detail on the matter, we note that we also
conducted further experiments in which the goal was not to benchmark
community detection algorithms, but rather to infer attribute values
as accurately as possible. In these experiments we included two
additional types of node features that one would almost certainly use
in practice. First, each node had features indicating its attributes
such gender, academic major, year of graduation, and dorm (of course,
we left out the attribute which we were trying to infer). Second, each
node had features indicating the percentage of its friends who had
each attribute value; e.g., the percentage of friends that were male
or female, and the percentage of friends in each possible academic
major. We found that these two simple feature types allowed for more
accurate classification than network communities. When we combined all
three feature types (i.e., each node's attributes, its friends
attributes, and the network communities to which it belonged), the
accuracy improved only slightly (by less than 1\%) over the case where
we left out the network communities altogether.

Thus, if one were trying to infer the missing attribute values in the
Facebook100 dataset, one could obtain quite good results even if one
ignored network communities altogether and use simpler features based
simply on node attributes the distribution of these attributes in each
node's egocentric network. While this may be a rather gloomy finding
for champions of community detection, we note that this finding does
not indicate the network communities are useless, but rather that in
the particular case of the Facebook100 dataset, these two other
features simply happen to contain information which is very useful for
inferring missing attribute values.

We conclude by noting that while the benchmarking here was based on
the task of inferring missing node attributes that we believe to be
closely related to community structure, one could construct
conceptually similar benchmarks based on different tasks. One natural
example would be to use network communities to perform supervised link
prediction ; this is a natural fit because presumable the processes
responsible for link formation are closely related to the processes
which form network communities.


\section*{Funding}
This work is supported by Science Foundation Ireland under grant
no. 08/SRC/I1407, Clique: Graph and Network Analysis Cluster.


\bibliographystyle{comnet}

\begin{thebibliography}{00}

\bibitem{Housing}
 (2004)  University of Chicago House system.
  \url{http://web.archive.org/web/20041031080505/http://www.rh.uchicago.edu/hds/housing/}.
Accessed: 28/12/2012.

\bibitem{Housing2}
 (2012)  Announcing New Residence Hall and Dining Commons.
  \url{http://csl.uchicago.edu/feature/announcing-new-residence-hall-and-dining-commons}.
Accessed: 28/12/2012.

\bibitem{Ahn2010}
Ahn, Y., Bagrow, J. {\&} Lehmann, S. (2010)  Link communities reveal multiscale
  complexity in networks. {\em Nature}, \textbf{466}(7307), 761--764.

\bibitem{Unknown1933}
Author, U. (1933)  Emotions Mapped by New Geography. {\em The New York Times}.

\bibitem{Blondel2008}
Blondel, V., Guillaume, J., Lambiotte, R. {\&} Lefebvre, E. (2008)  Fast
  unfolding of communities in large networks. {\em Journal of Statistical
  Mechanics: Theory and Experiment}, \textbf{2008}(10), P10008.

\bibitem{Clauset2007}
Clauset, A., Moore, C. {\&} Newman, M. (2007)  Structural inference of
  hierarchies in networks. {\em Statistical network analysis: models, issues,
  and new directions}, pages 1--13.

\bibitem{Danon2005}
Danon, L., Guilera, A.~D., Duch, J. {\&} Arenas, A. (2005)  {Comparing
  community structure identification}. {\em Journal of Statistical Mechanics:
  Theory and Experiment}, \textbf{2005}(9), P09008--09008.

\bibitem{Davis1941}
Davis, A., Gardner, B. {\&} Gardner, M. (1941) {\em Deep south}.
University of Chicago Press Chicago.

\bibitem{Delvenne2010}
Delvenne, J., Yaliraki, S. {\&} Barahona, M. (2010)  Stability of graph
  communities across time scales. {\em Proceedings of the National Academy of
  Sciences}, \textbf{107}(29), 12755--12760.

\bibitem{Duch2005}
Duch, J. {\&} Arenas, A. (2005)  Community detection in complex networks using
  extremal optimization. {\em Physical review E}, \textbf{72}(2), 027104.

\bibitem{Forsyth1946}
Forsyth, E. {\&} Katz, L. (1946)  A Matrix Approach to the Analysis of
  Sociometric Data: Preliminary Report. {\em Sociometry}, \textbf{9}(4), pp.
  340--347.

\bibitem{Fortunato2010}
Fortunato, S. (2010)  {Community detection in graphs}. {\em Physics Reports},
  \textbf{486}(3-5), 75--174.

\bibitem{Freeman03}
Freeman, L.~C. (2003)  Finding social groups: A meta-analysis of the southern
  women data. In {\em Dynamic Social Network Modeling and Analysis. The
  National Academies}, pages 39--97. Press.

\bibitem{Girvan2002}
Girvan, M. {\&} Newman, M. (2002)  Community structure in social and biological
  networks. {\em Proceedings of the National Academy of Sciences},
  \textbf{99}(12), 7821--7826.

\bibitem{Lancichinetti2010}
Lancichinetti, A., Kivel{\"a}, M., Saram{\"a}ki, J. {\&} Fortunato, S. (2010)
  Characterizing the community structure of complex networks. {\em PLoS One},
  \textbf{5}(8), e11976.

\bibitem{Lee2010}
Lee, C., Reid, F., McDaid, A. {\&} Hurley, N. (2010)  {Detecting highly
  overlapping community structure by greedy clique expansion}. {\em SNA-KDD
  2010}, page~11.

\bibitem{Marras2010}
Marras, E., Travaglione, A., Chaurasia, G., Futschik, M. {\&} Capobianco, E.
  (2010)  Inferring modules from human protein interactome classes. {\em BMC
  systems biology}, \textbf{4}(1), 102.

\bibitem{McDaid2011}
McDaid, A., Greene, D. {\&} Hurley, N. (2011)  Normalized Mutual Information to
  evaluate overlapping community finding algorithms. {\em arXiv preprint
  arXiv:1110.2515}.

\bibitem{Mislove2010}
Mislove, A., Viswanath, B., Gummadi, K. {\&} Druschel, P. (2010)  You are who
  you know: inferring user profiles in online social networks. In {\em
  Proceedings of the third ACM international conference on Web search and data
  mining}, pages 251--260. ACM.

\bibitem{Moreno1934}
Moreno, J. (1934) {\em Who shall survive? : a new approach to the problem of
  human interrelations}.
Nervous and Mental Disease Publishing Co.

\bibitem{Newman2006}
Newman, M. (2006)  Modularity and community structure in networks. {\em
  Proceedings of the National Academy of Sciences}, \textbf{103}(23),
  8577--8582.

\bibitem{scikit-learn}
Pedregosa, F., Varoquaux, G., Gramfort, A., Michel, V., Thirion, B., Grisel,
  O., Blondel, M., Prettenhofer, P., Weiss, R., Dubourg, V., Vanderplas, J.,
  Passos, A., Cournapeau, D., Brucher, M., Perrot, M. {\&} Duchesnay, E. (2011)
   {Scikit-learn: Machine Learning in Python }. {\em Journal of Machine
  Learning Research}, \textbf{12}, 2825--2830.

\bibitem{Rosvall2011}
Rosvall, M. {\&} Bergstrom, C. (2011)  Multilevel compression of random walks
  on networks reveals hierarchical organization in large integrated systems.
  {\em PloS one}, \textbf{6}(4), e18209.

\bibitem{Sampson1968}
Sampson, S. (1968) {\em A novitiate in a period of change: An experimental and
  case study of social relationships}.
PhD thesis, Cornell University.

\bibitem{Schaub2012}
Schaub, M., Delvenne, J., Yaliraki, S. {\&} Barahona, M. (2012)  Markov
  dynamics as a zooming lens for multiscale community detection: non
  clique-like communities and the field-of-view limit. {\em PloS one},
  \textbf{7}(2), e32210.

\bibitem{Stabeler2011}
Stabeler, M., Lee, C., Williamson, G. {\&} Cunningham, P. (2011)  Using
  Hierarchical Community Structure to Improve Community-Based Message Routing.
  In {\em ICWSM 2011 Workshop on Social Mobile Web Workshop, SMW}.

\bibitem{Traud2011}
Traud, A., Mucha, P. {\&} Porter, M. (2011)  Social structure of Facebook
  networks. {\em Physica A: Statistical Mechanics and its Applications}.

\bibitem{Traud2012}
Traud, A.~L., Mucha, P.~J. {\&} Porter, M, A. (2012)  {Social structure of
  Facebook networks}. {\em Physica A: Statistical Mechanics and its
  Applications}, \textbf{391}(16), 4165--4180.

\bibitem{Yang2012a}
Yang, J. {\&} Leskovec, J. (2012a)  Defining and evaluating network communities
  based on ground-truth. In {\em Proceedings of the ACM SIGKDD Workshop on
  Mining Data Semantics}, page~3. ACM.

\bibitem{Yang2012b}
Yang, J. {\&} Leskovec, J. (2012b)  Structure and Overlaps of Communities in
  Networks. In {\em Proceedings of the 6th SNA-KDD Workshop}.

\bibitem{Zachary1977}
Zachary, W.~W. (1977)  An Information Flow Model for Conflict and Fission in
  Small Groups. {\em Journal of Anthropological Research}, \textbf{33}(4), pp.
  452--473.

\end{thebibliography}

\end{document}